\documentclass[reprint,superscriptaddress,aps,pre]{revtex4-2}
\usepackage[usenames,dvipsnames]{xcolor}

\usepackage{natbib}
\usepackage{tikz}

\usepackage{float}

\usepackage{amsmath,amsfonts,amsthm,graphicx}
\usepackage{graphicx,psfrag}
\usepackage[psamsfonts]{amssymb}
\usepackage{amscd}
\usepackage[scr]{rsfso}
\usepackage[all,cmtip]{xy}
\usepackage{algorithm}
\usepackage{algorithmic}
\usepackage{adjustbox}
\usepackage{dsfont}
\usepackage{mathtools}
\usepackage{soul,xcolor}
\usepackage[colorlinks=true,linkcolor=Blue,citecolor=Blue,urlcolor=Blue]{hyperref}

\makeatletter
\def\BState{\State\hskip-\ALG@thistlm}
\makeatother

\usepackage{setspace}

\usepackage{dcolumn}
\usepackage{bm}
\usepackage[utf8]{inputenc}
\usepackage{accents}

\usepackage[caption=false]{subfig}
\usepackage{tikz}
\usetikzlibrary{arrows.meta}
\usepackage{array}
\usepackage[capitalise]{cleveref}

\setstcolor{magenta}

\theoremstyle{plain}

\begin{document}

\title{Forget Partitions: Cluster Synchronization in Directed Networks Generate Hierarchies}

\author{Fiona M. Brady}
\affiliation{Department of Physics and Astronomy, Northwestern University, Evanston, Illinois 60208, USA}
\author{Yuanzhao Zhang}
\affiliation{Department of Physics and Astronomy, Northwestern University, Evanston, Illinois 60208, USA}
\affiliation{Center for Applied Mathematics, Cornell University, Ithaca, New York 14853, USA}
\author{Adilson E. Motter}
\affiliation{Department of Physics and Astronomy, Northwestern University, Evanston, Illinois 60208, USA}
\affiliation{Northwestern Institute on Complex Systems, Northwestern University, Evanston, Illinois 60208, USA}

\begin{abstract} 
We present a scalable approach for simplifying the stability analysis of cluster synchronization patterns on directed networks. When a network has directional couplings, decomposition of the coupling matrix into independent blocks (which in turn decouples the variational equation) is no longer adequate to reveal the full relations among perturbation modes. Instead, it is often necessary to introduce directional dependencies among the blocks and establish hierarchies among perturbation modes. For this purpose, we develop an algorithm that finds the simultaneous block upper triangularization of sets of asymmetric matrices, which generalizes the Jordan canonical decomposition from a single matrix to an arbitrary number of matrices. The block upper triangularization orders subspaces of the variational equation in a directional manner, allowing the stability of perturbation modes to be analyzed in sequence. We show that our algorithm gives the greatest possible simplification under mild assumptions, both in terms of the sizes of the blocks and in terms of the number of nonzero upper triangular entries linking the blocks.
\end{abstract}

\maketitle

\section{Introduction}

Synchronization is one of the most fundamental forms of collective dynamics that can emerge from groups of interacting entities \cite{arenas2008synchronization}.
Maintaining a stable synchronized state is critical to the function of many real-world systems, such as power grids \cite{rohden2012self,dorfler2013synchronization,motter2013spontaneous} and circadian clocks \cite{aton2005vasoactive,to2007molecular,zhang2020energy}.
Importantly, global synchronization is not the only form of synchronization that matters, as complex networks often support complex synchronization patterns---two or more internally coherent but mutually independent clusters are known to often coexist in a single network \cite{kaneko1990clustering,belykh2001cluster,nakao2007noise,belykh2008cluster,skardal2011cluster,dahms2012cluster,rosin2013control,nicosia2013remote,fu2013topological,orosz2014decomposition,jalan2016cluster,zhang2017incoherence,hart2017experiments,menara2019stability,tang2019master,hart2019topological,zhang2020critical,zhang2021mechanism}.

Given a network of coupled oscillators, a fundamental question concerning any compatible synchronization pattern is its stability, since this determines whether the pattern can persist in the presence of unavoidable noise and perturbations.
With a few exceptions, previous studies on the stability of cluster synchronization patterns have focused exclusively on undirected networks \cite{pecora2014cluster,schaub2016graph,sorrentino2016complete,cho2017stable,siddique2018symmetry,zhang2020symmetry,zhang2020unified}.
This is in no small part due to the technical challenges imposed by directional couplings: directed networks introduce asymmetric matrices into the stability analysis, whose mathematical structure is significantly more complex than that of symmetric matrices \cite{lam2013first}.

\begin{figure*}[tb]
\centering
\subfloat[]{
\includegraphics[width=.85\linewidth]{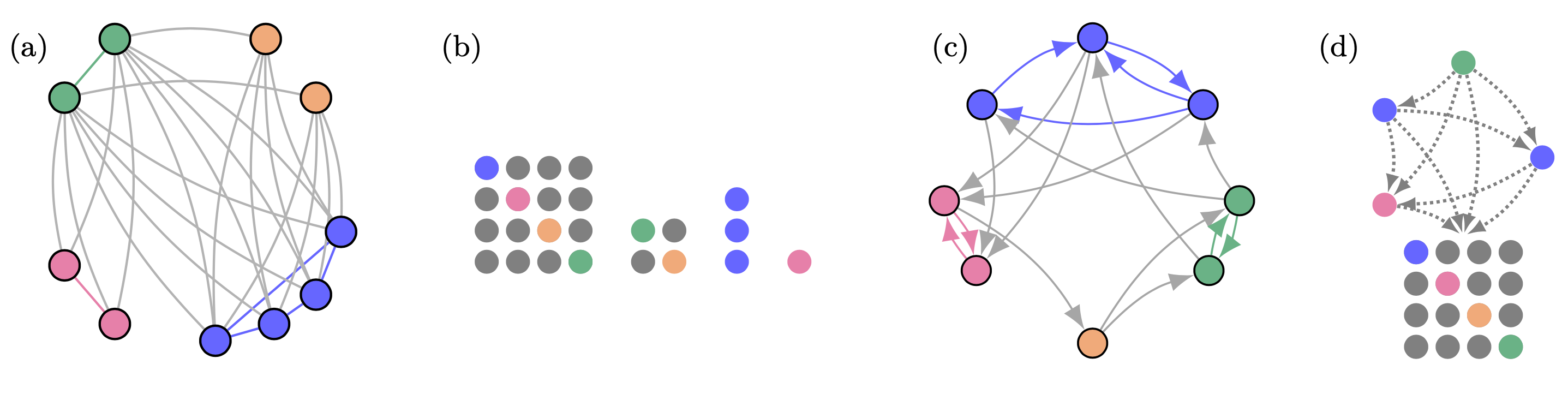}
}
\vspace{-5mm}
\caption{Cluster synchronization in directed networks generate hierarchies among perturbation modes. (a) Synchronization pattern on an undirected network. (b) Perturbation modes for the system in (a), which form independent blocks. (c) Synchronization pattern on a directed network. (d) Perturbation modes for the system in (c), which form a hierarchical structure.}
\label{fig:carton}
\end{figure*}

Due to this lack of proper theoretical tools, and despite its importance for the modeling of real-world systems \cite{newman2003structure,guimera2007module,timme2007revealing,leicht2008community,yu2011consensus,li2013epidemic,liu2016breakdown}, the treatment of directed networks in the context of cluster synchronization has been sporadic and limited to specific classes of synchronization patterns and coupling topologies \cite{lodi2020analyzing,salova2021cluster}.
Here, we fill this gap and develop a method to analyze the stability of arbitrary synchronization patterns in general networks with directional couplings.
Our approach is based on finding a simultaneous block upper triangularization of the (asymmetric) matrices in the variational equation, which we show to automatically reveal hierarchical dependencies that may exist among perturbation modes.
This is in stark contrast to the case of undirected networks, in which the perturbations can be decoupled and belong to independent blocks in the decomposed (symmetric) matrices.
This key difference between the directed and undirected cases is illustrated in \cref{fig:carton}.

This article is organized as follows.
\Cref{sec:bg} describes the class of systems we consider and outlines the challenges introduced by directed networks.
In \cref{sec:exp}, we illustrate the emergence of hierarchies among perturbations and discuss their implications to stability analysis using representative examples.
In \cref{sec:alg}, we present a simple algorithm for finding a simultaneous block upper triangularization (SBUT) of an arbitrary number of general matrices, which generalizes the simultaneous block diagonalization (SBD) technique developed for undirected networks \cite{murota2010numerical,maehara2011algorithm,irving2012synchronization,zhang2020symmetry,zhang2020unified} as well as the Jordan canonical form (developed for a single matrix).
Finally, we discuss potential extensions of our method in \cref{sec:end}.

\section{Cluster Synchronization in Directed Networks}
\label{sec:bg}

We consider networks of coupled oscillators described by
\begin{equation}  
 \dot{\bm{x}}_i = \bm{F}(\bm{x}_i) + \sigma \sum \limits_{j = 1}^n M_{ij} \bm{H}(\bm{x}_j), \quad i = 1,\dots,n,
\label{eq:dyn}
\end{equation}
where $\bm{x}_i$ is the $d$-dimensional state vector of the $i$th oscillator.
Here, $\bm{F}: \mathbb{R}^d \rightarrow \mathbb{R}^d$ describes how the individual oscillators would evolve if they were uncoupled, $\sigma$ is the coupling strength, $\bm{M}=(M_{ij})$ is the coupling matrix accounting for the network structure, and $\bm{H}$ is the interaction function.

For any given network system described by \cref{eq:dyn}, the compatible synchronization patterns can be derived from balanced equivalence relations, also called equitable partitions \cite{stewart2003symmetry,golubitsky2006nonlinear,belykh2011mesoscale,aguiar2011dynamics,kamei2013computation,aguiar2014lattice,golubitsky2016rigid,steur2016characterization,aguiar2018synchronization,nijholt2019center,Neuberger2020Invariant}, which ensures that oscillators in the same cluster can admit equal dynamics for generic $\bm{F}$ and $\bm{H}$. 
That is, an equitable partition defines an invariant synchrony subspace in which the cluster synchronization state is flow-invariant under the evolution of \cref{eq:dyn}.

To probe the stability of a compatible synchronization pattern, one needs to study the following variational equation, which describes the evolution of small deviations away from the cluster synchronization state:
\begin{equation}
    \delta\dot{\bm{X}} = \big( \sum_{k=1}^K \bm{E}^{(k)} \otimes \mathrm{J}\bm{F}(\bm{s}_k) + \sigma \sum_{k=1}^K \bm{M}\bm{E}^{(k)} \otimes \mathrm{J}\bm{H}(\bm{s}_k) \big) \delta\bm{X}.
  \label{eq:var-eq}
\end{equation}
Here, $\bm{s}_k$ is the synchronization trajectory of the $k$th cluster, $\delta\bm{X} = (\delta \bm{x}_1^\intercal, \cdots, \delta \bm{x}_n^\intercal)^\intercal$ is the $nd$-dimensional perturbation vector, and $\mathrm{J}$ is the Jacobian operator. 
Let $\mathcal{C}_k$ denote the set of nodes in the $k$th cluster, then
\[
  \bm{E}^{(k)}_{ii} =
  \begin{cases}
    1 & \quad \text{if } i \in \mathcal{C}_k \\
    0 & \quad \text{otherwise} \\
  \end{cases}
\]
is an $n \times n$ diagonal matrix encoding the nodes in the $k$th cluster.
For large $n$ typical of complex networks of scientific interest, the state space is high dimensional and a direct inspection of \cref{eq:var-eq} offers limited insight into the stability of cluster synchronous states.

The challenge to simplify \cref{eq:var-eq} is that coordinates that simplify matrix $\bm{M}$ can make the matrices $\bm{E}^{(k)}$ more complex, and vice versa. The ideal approach is thus to identify coordinates that simplify both terms simultaneously to the maximum extent possible.
When the network is undirected, $\{\bm{E}^{(k)}\}$ and $\bm{M}$ are all symmetric matrices, and \cref{eq:var-eq} can be decoupled by simultaneously decomposing these matrices into independent blocks of minimal sizes.
This decomposition can be achieved by an orthogonal transformation found through either group theoretic \cite{pecora2014cluster,sorrentino2016complete} or algebraic techniques \cite{zhang2020symmetry,zhang2020unified,salova2021cluster}.
However, when the network is directed, the matrix $\bm{M}$ becomes asymmetric.
We know that even for a single asymmetric matrix, orthogonal transformations are often unable to find its optimal decomposition, which takes the Jordan canonical form \cite{meyer2000matrix,lu2006new,nishikawa2006maximum}.
Thus, in order to analyze \cref{eq:var-eq} for directed networks, we need to consider general similarity transformations and accommodate directional dependencies among blocks.
These directional dependencies are encoded by nonzero upper triangular entries in the transformed matrices, and have significant implications for stability analysis, as we show below.

\begin{figure*}[tb]
\centering
\subfloat[]{
\includegraphics[width=.85\linewidth]{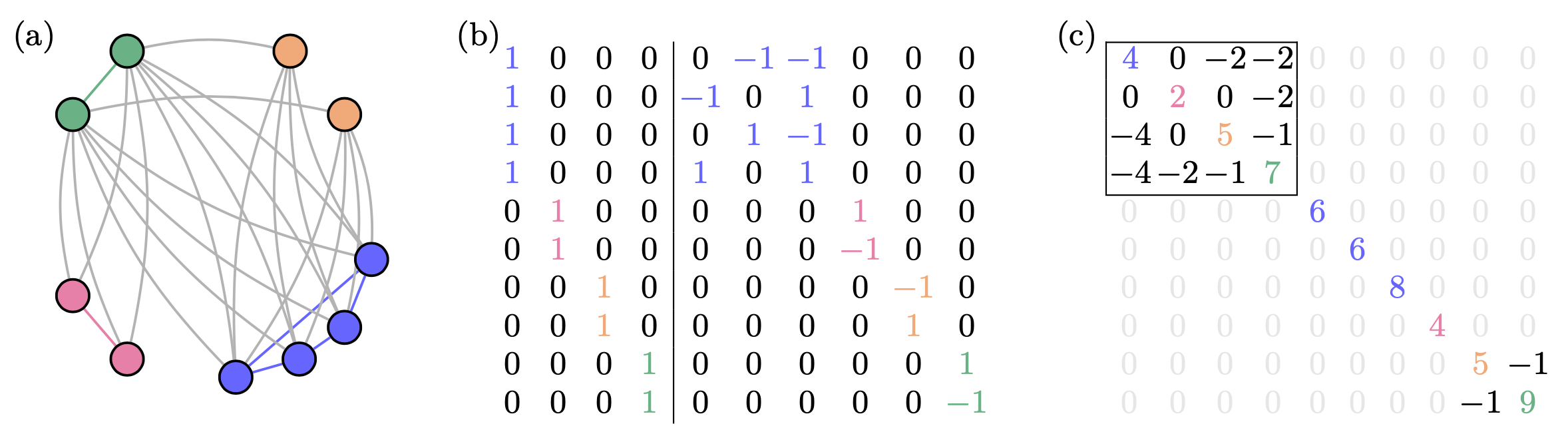}
}
\vspace{-5mm}
\caption{Example undirected network with ten nodes and four clusters. (a) Network diagram, where nodes and intracluster connections are colored by clusters. (b) Basis found by the SBUT algorithm. The nonzero entries are color-coded to match the corresponding clusters. The first four columns represent perturbations parallel to the cluster synchronization manifold, and the rest of the columns encode perturbations transverse to the cluster synchronization manifold. (c) Graph Laplacian under the SBUT coordinates. The first $4 \times 4$ block (marked by the black box) corresponds to parallel perturbations and is excluded from the stability analysis.}
\label{fig:ex_1}
\end{figure*}

\begin{figure*}[tb]
\centering
\subfloat[]{
\includegraphics[width=.65\linewidth]{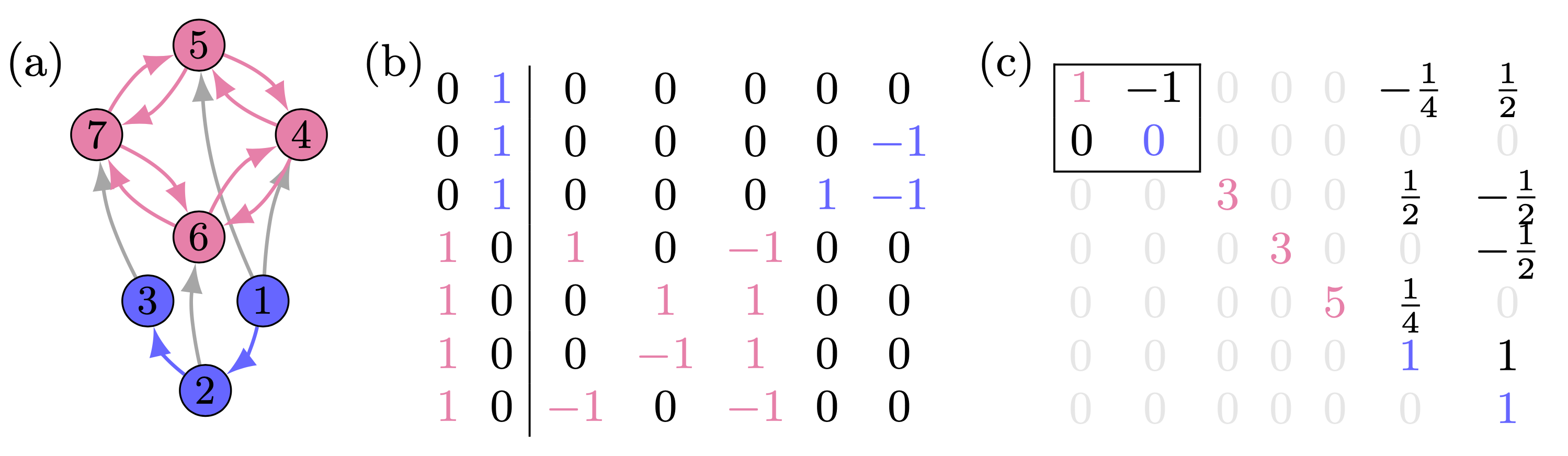}
}
\vspace{-5mm}
\caption{Example directed network composed of a master cluster and a slave cluster. (a--c) Network diagram (a), transformation matrix found by the SBUT algorithm (b), and graph Laplacian under the SBUT coordinates (c). Unlike the case of undirected networks, there are directional dependencies between different blocks, as evidenced by the nonzero upper triangular entries in the transformed graph Laplacian. Specifically, perturbations inside the master cluster (colored blue) influence perturbations inside the slave cluster (colored pink), but not the other way around.}
\label{fig:ex_2}
\end{figure*}

\begin{figure*}[tb]
\centering
\subfloat[]{
\includegraphics[width=.7\linewidth]{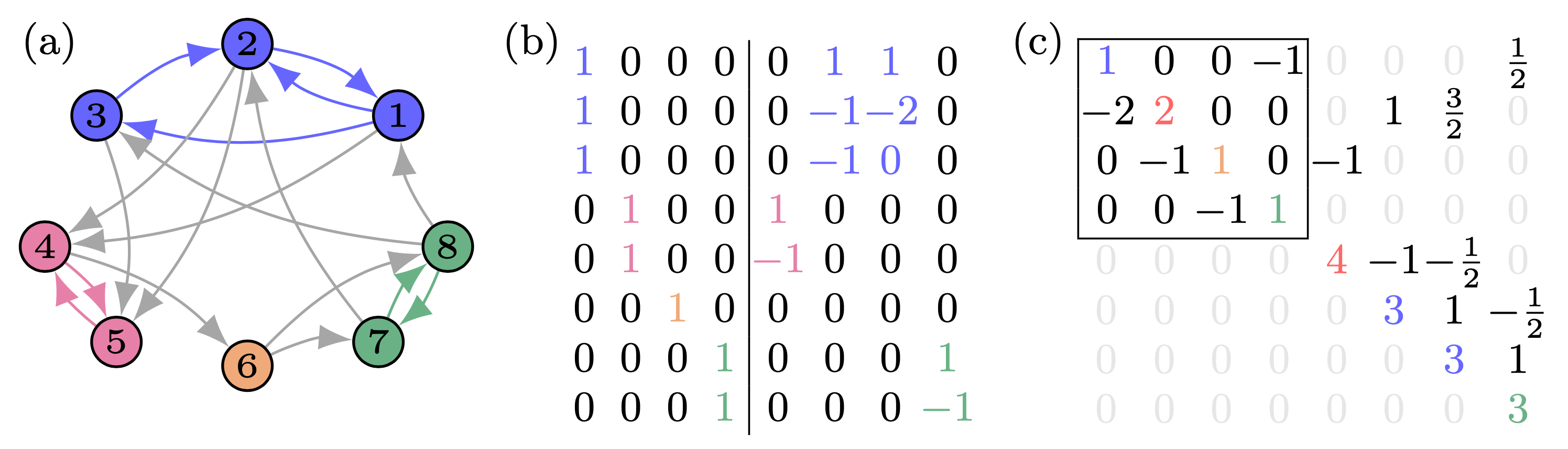}
}
\vspace{-5mm}
\caption{Example directed network with circular flow among clusters (blue$\rightarrow$pink$\rightarrow$orange$\rightarrow$green$\rightarrow$blue). (a--c) Network diagram (a), transformation matrix found by the SBUT algorithm (b), and graph Laplacian under the SBUT coordinates (c). Even in the absence of a master-slave structure among clusters, SBUT revels a clear hierarchy in the perturbation space. Note that SBD can only find a trivial $8\times 8$ block for this network and would have missed the intricate dependencies among perturbation modes.}
\label{fig:ex_3}
\end{figure*}

\begin{figure*}[tb]
\centering
\subfloat[]{
\includegraphics[width=.65\linewidth]{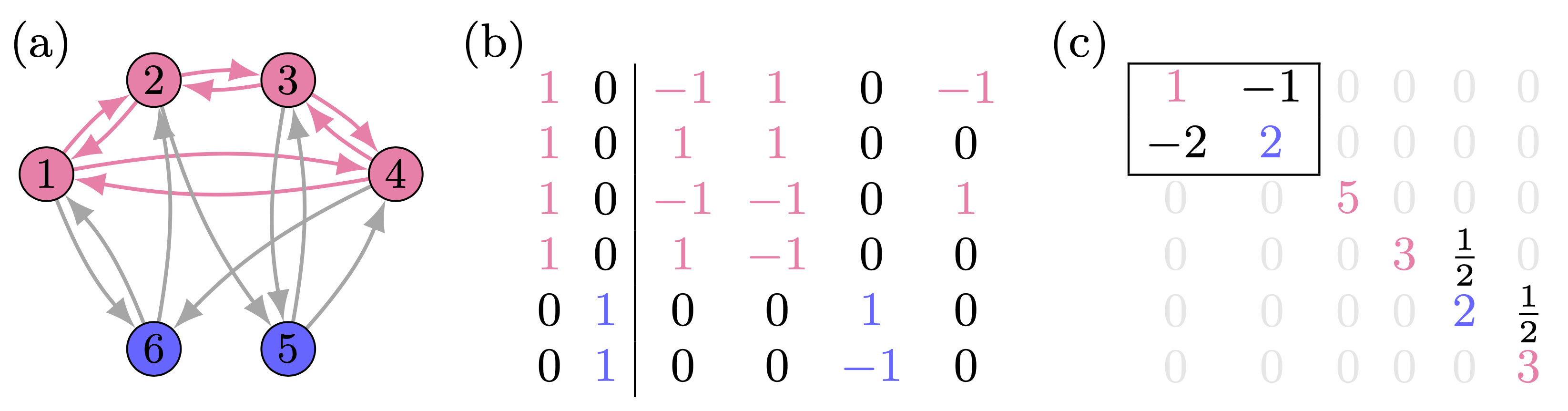}
}
\vspace{-5mm}
\caption{Example directed network composed of intertwined clusters with bidirectional flow. (a--c) Network diagram (a), transformation matrix found by the SBUT algorithm (b), and graph Laplacian under the SBUT coordinates (c). Interestingly, the transverse perturbation inside the blue cluster, $(0,0,0,0,1,-1)^\intercal$, is sandwiched between two transverse perturbation modes from the pink cluster. Moreover, $(-1,0,1,0,0,0)^\intercal$ only influences $(1,1,-1,-1,0,0)^\intercal$ through $(0,0,0,0,1,-1)^\intercal$, which is from the other cluster. This shows that intertwined clusters in directed networks is a rich concept, and many different levels of intertwinedness is possible.}
\label{fig:ex_4}
\end{figure*}

\begin{figure}[tb]
\includegraphics[width=\columnwidth]{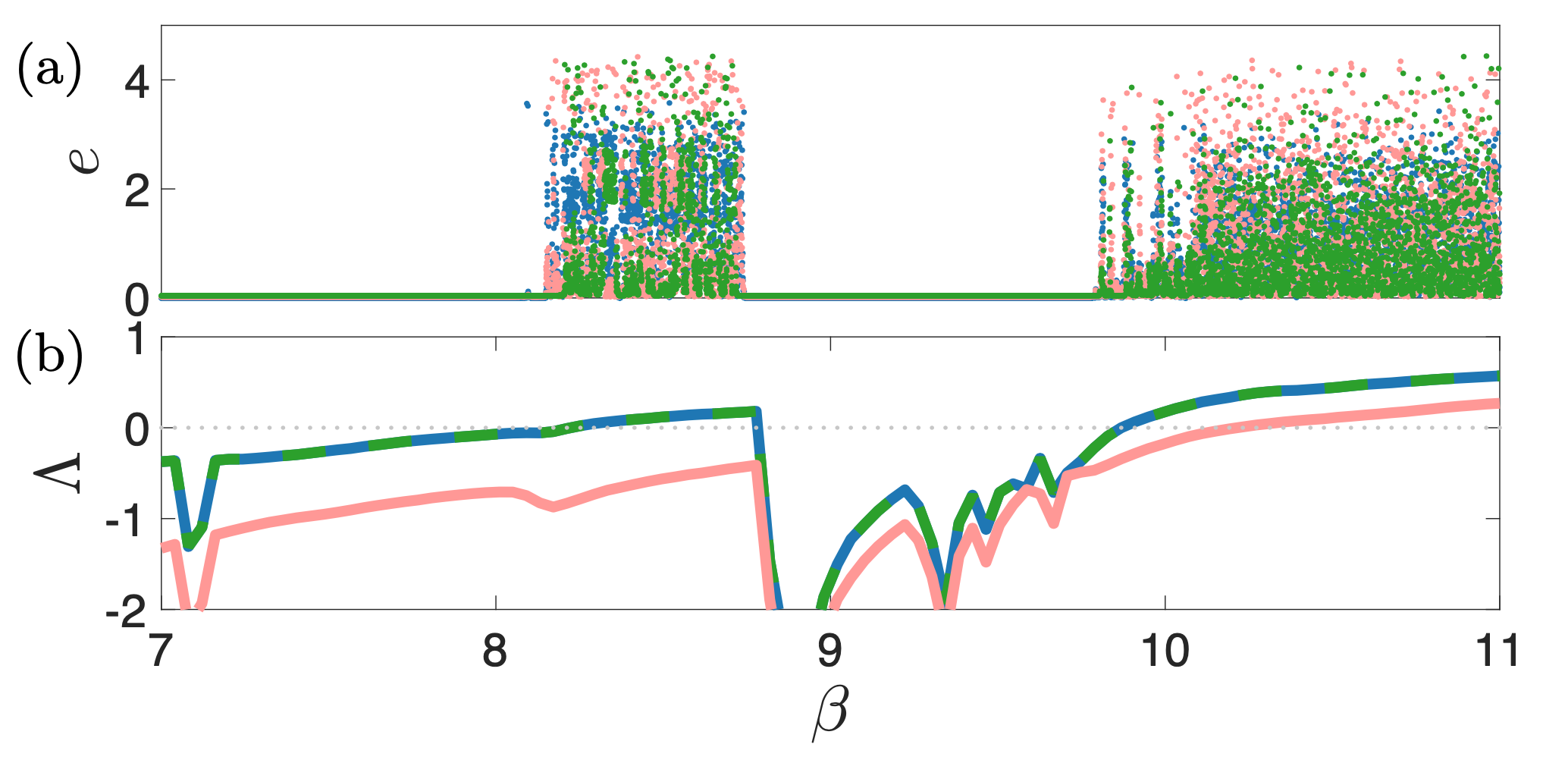}
\caption{Stability analysis of the synchronization pattern in \cref{fig:ex_3} based on SBUT coordinates. Each node models an optoelectronic oscillator with self-feedback strength $\beta$. (a) Synchronization error $e$ in the three clusters with two or more nodes as $\beta$ is quasistatically decreased from $11$ to $7$. The color coding matches that of \cref{fig:ex_3}. (b) Lyapunov exponents $\Lambda$ of the transverse perturbation modes calculated from SBUT coordinates. Although the perturbation mode corresponding to the pink cluster by itself is stable for $\beta < 10.2$, it is influenced by the other transverse perturbation modes that are unstable for a wider range of $\beta$. Thus, all three clusters (including the pink cluster) lose synchrony concurrently when the blue/green curve in (b) crosses zero from below.}
\label{fig:ex_3_stab}
\end{figure}

\section{Hierarchies in the Perturbation Space}
\label{sec:exp}

Before describing in detail how to find SBUT coordinates, we first demonstrate the utility of our approach for stability analysis and point out key structures that only emerge in the case of directed networks.
For this purpose, we consider four representative examples with increasing complexity and in each case we discuss the SBUT structure identified by the algorithm we introduce in \cref{sec:alg}.
In these examples, we assume the couplings to be diffusive, which is typically more challenging to deal with than non-diffusive couplings \cite{sorrentino2016complete}.
Thus, the coupling matrices take the form of graph Laplacians $\bm{L} = \bm{D} - \bm{A}$, where $\bm{A}$ is the adjacency matrix and $\bm{D}$ is a diagonal matrix whose entries are given by the rowsums of $\bm{A}$.

The first example is a demonstration that SBUT reduces to SBD in the case of undirected networks. 
The network shown in \cref{fig:ex_1} is an undirected network with ten nodes and four clusters (nodes in the same cluster share the same color).
By applying the SBUT algorithm, we find a basis for SBUT coordinates depicted in \cref{fig:ex_1}(b).
The graph Laplacian under the SBUT coordinates is in a block diagonal form, as expected for an undirected network (and symmetric matrices).
From the last diagonal block, we know that the orange and the green clusters are intertwined \cite{pecora2014cluster}.

In the second example, shown in \cref{fig:ex_2}, the network is formed by two clusters, with intercluster coupling only going from the blue cluster to the pink cluster.
Already in this simple example, we see some key differences from the case of undirected networks.
First, for directed networks, perturbations parallel to the cluster synchronization manifold can be influenced by those transverse to the manifold, whereas for undirected networks the parallel and the transverse perturbations belong to independent blocks.
Second, there are directional dependencies among perturbation modes (one mode can influence another mode, but not the other way around).
Both features above are reflected by the nonzero upper triangular entries in the transformed graph Laplacian [\cref{fig:ex_2}(c)].
This block upper triangular form significantly simplifies the stability analysis.
The $2 \times 2$ block in the upper left corresponds to parallel perturbations and is excluded from the stability analysis. 
The analysis therefore boils down to finding the Lyapunov exponents of the five remaining $1 \times 1$ blocks.
This calculation can be done sequentially, starting from the blue block in the lower right corner.
Once we have encountered an unstable block, the analysis can stop because all blocks ``downstream'' to the unstable block are automatically unstable.
On the other hand, a block diagonal form cannot achieve meaningful decomposition in this case and already fails at separating the parallel perturbations from the transverse ones.

In \cref{fig:ex_3}, we consider a network consisting of four clusters, with a circular flow among the clusters. 
Again, the block triangular form reveals the hierarchy among the perturbation modes.
We note that for finding a finest block diagonal decomposition, orthogonal basis is always sufficient \cite{zhang2020symmetry}.
However, in order to reach an optimal block upper triangular form, non-orthogonal basis is often necessary.
In the current example, the SBUT coordinates shown in \cref{fig:ex_3}(b) contain two base vectors, $(1,-1,-1,0,0,0,0,0)^\intercal$ and $(1,-2,0,0,0,0,0,0)^\intercal$, that are not orthogonal to each other.

Finally, in \cref{fig:ex_4}, we consider the most general case---directed networks with bidirectional flow between intertwined clusters.
From \cref{fig:ex_4}(c), we see that the two clusters are intertwined in an intricate manner.
Aside from the perturbation mode $(-1,1,-1,1,0,0)^\intercal$, which forms its own invariant subspace, the three other transverse modes form a hierarchy, with the mode from the blue cluster sandwiched between two modes from the pink cluster.
From this structure, we know that loss of synchrony in the blue cluster will inevitably desynchronize the pink cluster, whereas loss of synchrony in the pink cluster can either desynchronize the blue cluster or not depending on which mode is unstable. 
As a result, coherence and incoherence can coexist in this network despite the fact that the two clusters are mutually intertwined.

To show a concrete example of how SBUT coordinates can be used in stability analysis, we study the synchronization pattern in \cref{fig:ex_3} for node dynamics that model optoelectronic oscillators \cite{pecora2014cluster}.
The system dynamics is described by the following equations:
\begin{equation}
  x_i^{t+1} = \beta [1-\cos(x_i^t)]/2 - \sigma \sum_{j=1}^{n} L_{ij} [1-\cos(x_j^t)]/2 + \Delta, 
\label{eq:opto}
\end{equation}
where we set the coupling strength to $\sigma=1.5$ and the offset parameter to $\Delta=0.525$.
We also constrain the dynamical variables $x_i$ to the interval $[0,2\pi)$ by taking $\text{mod} \, 2\pi$ at each iteration.
In \cref{fig:ex_3_stab}(a), we quasistatically decrease the self-feedback strength $\beta$ from $11$ to $7$ and monitor the synchronization error $e_k = \sqrt{\sum_{i\in \mathcal{C}_k} (x_i - \bar{x})^2/n_k}$ in the three clusters with $n_k>1$ node.
It is observed that the three clusters are either all synchronized or all desynchronized.

From \cref{fig:ex_3}(c), we know the perturbation mode localized in the green cluster controls those in the blue cluster, which in turn control the one in the pink cluster.
As a consequence, we can analyze the last four $1\times 1$ diagonal blocks in \cref{fig:ex_3}(c) sequentially (each corresponding to a perturbation mode).
If the green block is unstable, then the blue and the pink blocks are automatically unstable.
If the green block is stable, then we can ignore the upper triangular entry linking the green block to the next blue block and analyze the stability of the blue block as if it was an independent block.
\Cref{fig:ex_3_stab}(b) shows the Lyapunov exponents $\Lambda$ for the green, blue, and pink blocks when analyzed as independent blocks.
Since the pink block is ``downstream'' from the green and the blue blocks, we know that it will inherit instability from them.
Thus, although $\Lambda<0$ for the pink block when $\beta<10.2$, the pink cluster still loses synchrony alongside the green and the blue clusters (e.g., for $8.2<\beta<8.7$).

\section{Simultaneous Block Upper Triangularization of General Matrices}
\label{sec:alg}

We now present a practical algorithm to find a similarity transformation matrix $\bm{T}$ that simultaneously block upper triangularizes multiple matrices.
Given a set of matrices $\mathcal{B} = \{\bm{B}^{(1)},\bm{B}^{(2)},\dots,\bm{B}^{(\mathscr{L})}\}$, the algorithm consists of three simple steps:
\begin{enumerate}
  \item Find the generalized eigenvectors $\bm{v}_i$ of the matrix $\bm{B} = \sum_{\ell=1}^\mathscr{L} \xi_\ell \bm{B}^{(\ell)}$, where $\xi_\ell$ are independent random coefficients drawn from a Gaussian distribution. Set $\bm{T} = [\bm{v}_1,\cdots,\bm{v}_n]$.
  \item Generate $\bm{B}^\prime = \sum_{\ell=1}^\mathscr{L} \xi_\ell^\prime \bm{B}^{(\ell)}$ for a new realization of $\xi_\ell^\prime$ and compute $\widetilde{\bm{B}}=\bm{T}^{-1} \bm{B}^\prime \bm{T}$.
  \item Set $\bm{T} = [\bm{v}_{\epsilon(1)},\cdots,\bm{v}_{\epsilon(n)}]$, where $\epsilon$ is a permutation of $1,\cdots,n$ such that if $\widetilde{B}_{ij}\neq0$ and $\widetilde{B}_{ji}=0$ then $\bm{v}_i$ is ordered before $\bm{v}_j$. 
\end{enumerate}

We say a block upper triangular form is {\it finest} when it maximizes the number of diagonal blocks.
On top of that, we also require that $\bm{T}$ minimizes the number of nonzero upper triangular entries linking those finest blocks.
The latter condition is important to avoid spurious dependencies between blocks, which could for example lead to the incorrect conclusion that a perturbation mode is unstable due to the falsely-inferred influence of a different mode. 
We show next that as long as the random matrix $\bm{B}$ is not derogatory (i.e., no two of its eigenvectors have identical eigenvalues), the similarity transformation $\bm{T}$ found by the algorithm above is guaranteed to generate a finest SBUT of the matrix set $\mathcal{B}$.
(For a more detailed argument, see \cref{sec:proof}).

In the absence of degeneracy, since there is no ambiguity in choosing eigenvectors, the Jordan canonical form encodes all the subspaces of $\mathbb{R}^n$ that are invariant under the action of $\bm{B}$ and captures their inclusion relations.
On the other hand, the finest SBUT form reflects all the subspaces of $\mathbb{R}^n$ that are invariant under the action of the matrix set $\mathcal{B} = \{\bm{B}^{(1)},\bm{B}^{(2)},\dots,\bm{B}^{(\mathscr{L})}\}$. 
These subspaces are a subset of the invariant subspaces under the action of the single matrix $\bm{B}$.
Since the generalized eigenvectors $\{\bm{v}_i\}$ capture all the invariant subspaces of $\bm{B}$, they also capture its subset that are invariant under the action of $\mathcal{B}$.
Thus, $\{\bm{v}_i\}$ form a basis that gives the finest SBUT of $\mathcal{B}$ when properly ordered.

In terms of the computations, our algorithm reduces the task of finding the SBUT of multiple matrices to finding the Jordan canonical form of a {\it single} matrix, which can be carried out with standard software packages \cite{meurer2017sympy}.
The Jordan decomposition is the most computationally intensive part of the algorithm and scales as $\mathcal{O}(n^3)$ \cite{kaagstrom1980algorithm,beelen1988improved}.
A Python and Mathematica implementation of the SBUT algorithm is available online as part of this publication \footnote{\url{https://github.com/y-z-zhang/SBUT}}.

We note that for applications to cluster synchronization, due to the special structure of $\{\bm{E}^{(k)}\}$, it is often more convenient to consider $\bm{B}=\sum_{k=1}^K \xi_k \bm{E}^{(k)}\bm{M}\bm{E}^{(k)}$ instead of a random linear combination of $\{\bm{E}^{(k)}\}$ and $\bm{M}$ in Step $1$ of the algorithm.
This choice of $\bm{B}$ automatically enforces the cluster structure in the transformation matrix $\bm{T}$, and each base vector $\bm{v}_i$ is localized within exactly one cluster.
As a bonus, the matrices $\{\bm{E}^{(k)}\}$ remain invariant under the transformation $\bm{T}$, so one only needs to look at $\bm{M}$ to identify the SBUT structure.

\section{Conclusion}
\label{sec:end}

Our study of synchronization patterns in directed networks  offers two main contributions. 
First, it shows that stability analysis is optimally simplified by a generally non-orthogonal transformation of the variational equation into a block upper triangular form. 
This, in turn, offers insight into the dynamical organization of a large class of networks of coupled oscillators. 
Second, it provides a simple yet broadly applicable numerical method for SBUT that is both effective and efficient, which enables the approach to be applied to large complex networks.

For the ease of presentation, we focused exclusively on directed networks with pairwise and time-independent coupling interactions of a single type.
However, the technique developed here can be easily extended to generalized networks, including hypergraphs \cite{berge1973graphs,battiston2020networks}, multilayer networks \cite{kivela2014multilayer,boccaletti2014structure}, and temporal networks \cite{holme2012temporal}.
In such cases, additional matrices appear in the stability analysis \cite{zhang2020unified}, and we anticipate that they can be accounted for naturally within the SBUT framework.
Applications of our method to directed networks with generalized interactions is thus an exciting direction for future research.

\section{Acknowledgements}

The authors thank Istv\'an Kov\'acs and Duan Chao for insightful discussions. This work was supported by ARO Grant No.\ W911NF-19-1-0383, a Schmidt Science Fellowship, and a Summer Research Grant from Northwestern's Weinberg School of Arts and Sciences (Summer 2019).

F.M.B. and Y.Z. contributed equally to this work.

\vspace{5mm}

\noindent {\bf Note:} After this work was completed, a related manuscript was made public almost at the same time as ours. In that work \cite{lodi2021one}, Lodi {\it et al.}\ tackled the stability problem of cluster synchronization in directed networks using a very different technique, leveraging the concepts of invariant synchrony subspaces and breaking vectors. It would be interesting to compare and potentially combine the techniques presented here and in Lodi {\it et al.}\ to gain further understanding of complex synchronization patterns in directed networks.

\appendix

\section{Optimality of the SBUT structure}
\label{sec:proof}

The Jordan basis for a matrix is a collection of eigenvectors, which are scaled by the matrix, and, in the case of nontrivial blocks, generalized eigenvectors, which are scaled and rotated. Importantly, for a matrix without degeneracy, such as $\bm{B}$, the matrix operates in a unique way on each (generalized) eigenvector. Using this asymmetry, it is possible to isolate these directions, using only sums and scalar multiples of repeated applications of $\bm{B}$, and to create projection operators onto the Jordan blocks. For nontrivial blocks, both projection onto the Jordan blocks and the rotations within them can be isolated. These projection operators are not orthogonal projections in general. They map other (generalized) eigenvectors to zero, and are only orthogonal when the basis is orthonormal. Relating these projections to the finest SBUT of the matrix set $\mathcal{B}$ provides crucial insight into how the SBUT is structured, because any basis which gives a SBUT of $\mathcal{B}$ also puts the projections into the same upper triangular form.

To further elucidate properties of the SBUT, it is useful to consider subspaces of the perturbation space that are closed with respect to $\mathcal{B}$. That is, when applying a matrix in $\mathcal{B}$ to any vector in the subspace, the result will also be a vector in the subspace. (Being a subspace means that the sum of any two vectors in the set remains in the set.) Note that the construction ensures that if a subspace is closed with respect to $\mathcal{B}$, applying the projection operators will also send vectors from the subspace to vectors in the same subspace. The important observation connecting these types of subspaces to a SBUT is that any SBUT corresponds to an increasing collection of such closed subspaces, namely the sequence where the spanning set of the $k$th subspace is the basis vectors for the first $k$ blocks.

For a subspace closed with respect to $\mathcal{B}$, any hypothetical basis can be re-expressed in terms of (generalized) eigenvectors of $\bm{B}$. As the projections map a basis for the subspace to those (generalized) eigenvectors of $\bm{B}$ which they are formed from, those (generalized) eigenvectors are contained in the closed subspace. Hence, it is possible to form a basis for any closed subspace in terms of only (generalized) eigenvectors. This shows that the Jordan basis of $\bm{T}$ will in fact give a SBUT of $\mathcal{B}$ with the smallest block sizes possible.

Further, it is fairly simple to show that this SBUT form has no unnecessary links between blocks. Two blocks in a SBUT are unconnected precisely when the smallest closed subspace containing the basis vectors for one block has a spanning set that can be expressed without reference to the basis vectors of the other block (and vice versa). All closed subspaces are spanned by (generalized) eigenvectors of $\bm{B}$, and are required to express elements of closed subspaces precisely when the basis vectors are contained in the closed subspaces. Choosing those (generalized) eigenvectors as the basis elements will eliminate unnecessary connections. Therefore the SBUT this basis gives is in fact the finest SBUT for $\mathcal{B}$.

This argument relies solely on the non-degeneracy of $\bm{B}$. The Jordan basis for any matrix derived from $\mathcal{B}$ with non-degenerate eigenspaces will give a way to find the finest SBUT for the matrix set. Choosing a generic linear combination will avoid unnecessary degeneracies. With probability $1$ a second generic element, $\widetilde{\bm{B}}$, represented in this basis will have nonzero entries everywhere any element of $\mathcal{B}$ has nonzero entries. Therefore, ordering the basis so that $\widetilde{\bm{B}}$ is in block upper triangular form will put all elements of $\mathcal{B}$ into such a form.

\bibliography{net_dyn,net_dyn_2}

\end{document}